
\input harvmac.tex

\def \Pp{{\bf P}}
\def \P1{${\bf P}^1$}
\def \Z2{${\bf Z}_2$}

\noblackbox
\Title{\vbox{\hbox{HUTP-95/A047, IASSNS-HEP-95/105}
\hbox{\tt hep-th/9511222}}}
{D-Branes and Topological Field Theories}
\bigskip\vskip2ex
\centerline{M. Bershadsky and C. Vafa}
\vskip2ex
\centerline{\it  Lyman Laboratory of Physics, Harvard
University}
\centerline{\it Cambridge, MA 02138, USA}
\vskip2ex
\centerline{and}
\vskip2ex
\centerline{V. Sadov}
\vskip2ex
\centerline{\it School of Natural Sciences and  School of
Mathematics, }
\centerline{\it Institute for Advanced Study}
\centerline{\it Olden Lane, Princeton, NJ 08540, USA}
\vskip .3in

In the presence of a D-brane a string theory develops
a new subsector.  We show that for curved D-branes the corresponding
sector is a (partially twisted) topological field theory.
We use this result to compute the degeneracy of 2-branes wrapped
around $K3$ cycles as well as 3-branes wrapped around CY threefold
vanishing 3-cycles.  In both cases we find the degeneracy is in accord
with expectation.  The counting of BPS states of a gas of
0-branes in the presence of a 4-brane
in $K3$ is considered and it is noted that the
effective 0-brane charge is shifted by 1, due to a quantum correction.
This is in accord with string duality and the fact
that left-moving ground state energy of heterotic string starts at $-1$.
We also show that all the three different topological
twistings of four dimensional $N=4$ Yang-Mills theory do arise from
curved three-branes embedded in different spaces
(Calabi-Yau manifolds and manifolds with exceptional holonomy groups).

 \Date{\it {Nov. 1995}}
%

\newsec{Introduction}
A significant progress has been made in understanding type II
string solitons by identifying the sources of Ramond-Ramond charges
as D-branes \ref\pol{J. Polchinski,
{\it Dirichlet-Branes and Ramond-Ramond Charges}, NSF-ITP-95-122,
hep-th/9510017}.
As specific applications this has led to verification
\ref\witb{E. Witten,
{\it Bound States Of Strings And $p$-Branes},
 hep-th/9510135 }\
of the
prediction of $(n,m)$ strings
as bound states as required by string duality \ref\sch{J. Schwarz,
{\it An SL(2,Z) Multiplet of Type IIB Superstrings}, CALT-68-2013,
hep-th/9508143}\ and connecting type
IIB $SL(2,{\bf Z})$ duality in ten dimensions
 to the appearance of massless gauge particles and matter
in type II compactifications on singular $K3$ and Calabi-Yau
threefolds \ref\bsv{M. Bershadsky, V. Sadov and C.Vafa,
{\it D-Strings on D-Manifolds}, HUTP-95/A035, IASSNS-HEP-95-77,
hep-th/9510225}
\ref\ov{Ooguri and C. Vafa.,
{\it Two-Dimensional Black Hole and Singularities of CY Manifolds},
hep-th/9511164}.  In addition a systematic
count of some expected $D$-brane bound states on $T^4$
\ref\senone{A. Sen,
{\it A Note on Marginally Stable Bound States in Type II String Theory},
MRI/PHY/23-95, hep-th/95102229}
\ref\sentwo{A. Sen,
{\it U-duality and Intersecting D-branes}, MRI/PHY/23-95,
hep-th/9511026}\
 has become possible and has led to counting
of some of the Hagedorn density of BPS states as ground state
configurations for a supersymmetric gas of D-branes \ref\va{C. Vafa,
{\it Gas of D-Branes and Hagedorn Density of BPS States}, HUTP-95/A042,
hep-th/9511088}.

In most of the above applications, the world volume of the D-brane
is flat.  One aim of this paper is to study the setup for extending
the above cases to more general cases where the world volume of the
D-brane is curved.  That this should be considered is natural
when we recall that in the case of conifold on $K3$ the
enhanced gauge symmetry expected at those points
\ref\witd{E. Witten, Nucl. Phys. {\bf B443} (1995) 85}\
comes from a 2-brane wrapped around
a vanishing $S^2$ and for the conifold point of a Calabi-Yau threefold,
the massless charged hypermultiplet comes from a 3-brane wrapped
around a vanishing $S^3$ \ref\strom{A. Strominger, Nucl. Phys. {\bf B451}
(1995) 96}.

Quite surprisingly we find that the new sector living on the
worldbrane is {\it automatically} topologically twisted
along the directions where the worldvolume is curved.
This is a very natural and satisfying realization of topological
field theories in the context of string theory.
There is, however,
a minor twist to the story:  The bosonic fields which are
no longer spin-0 due to twisting are partially compactified
carrying the information about the embedding of the D-brane
in the rest of the manifold.

The organization of this paper is as follows: In section
2 we discuss some generalities about the field theories
on curved D-brane worldvolumes
and their connection to topological field theories.
In section 3
we consider the specific case of type II compactifications on $K3$
and $T^4$.  In particular we show that the degeneracy
of the 2-branes on $K3$ wrapped around a vanishing
2-cycle of $K3$
 is 1, as expected.
Moreover we show that there are no bound states corresponding
to multiple wrappings.  We also show that 2-branes wrapped around
higher genus Riemann surfaces $\Sigma$ in $K3$ are a natural
candidate for the D-brane states corresponding to Hagedorn
density of BPS states.  We explicitly show this for genus one
curves in $K3$ (where genus $g$ corresponds to bosonic
oscillator number $N_L=g$).
In these cases we show
that the relevant topological theory on $\Sigma \times {\bf R}\times S^1$
 is of the form studied in \ref\bjsv{M. Bershadsky, A. Johansen, V.
Sadov and
C. Vafa,
Nucl. Phys. {\bf B448} (1995) 166}\ where
we have a partially twisted
(along $\Sigma$) $N=4$ theory.
In section 4 we discuss the counting of bound states of
$n$ 0-branes and one 4-brane on $K3$ and show how it leads
to the expected degeneracy on the heterotic side as
suggested in \va .  A key feature of this correspondence
is the appearance of an ``anomaly'' which causes a 4-brane
to carry $-1$ unit of 0-brane charge, thus shifting
the effective value of 0-branes $n\rightarrow n-1$.

In section 5 we discuss the counting for D-brane
states for the Calabi-Yau conifold and we again
find agreement with the prediction of the expected degeneracy \strom .
In section 6 we give other examples including manifolds of exceptional
holonomy $G_2$ and $Spin(7)$ and Calabi-Yau fourfolds and discuss
the implications of potential singularities for such cases.  We find that all
the three different twistings of $N=4$ topological theories
are realized by this set of examples.

\newsec{D-branes and Topological Field Theories}

In this section we sketch the general connection between
the D-branes and the topological field theories.  In the presence
of a D-brane, string theory develops a new subsector
of open strings. The most intensively studied case corresponds
to when the D-branes worldvolume is flat, in which case one
finds that the corresponding field theory is the
reduction of the $N=1$ Yang-Mills theory from 10 dimensions
to $p+1$, where we are considering p-branes.  However
more general curved worldbranes can be considered.
To see that this can be done, it suffices to recall how
the D-branes were originally discovered by
considering orientifolds
\ref\hor{P. Horava, Phys. Lett. {\bf B231} (1989) 251}
\ref\poldai{J. Dai, R. G. Leigh and J. Polchinski, Mod. Phys. lett,
{\bf A4}
(1989) 2073}:
  One considers modding out, in an oriented closed string theory, by a
reflection in spacetime, which at the same time acts on the
closed string by reversing its orientation.  Suppose such
a transformation has a fixed subspace in spacetime.  Then
we can consider open string states which end at these fixed
subspaces as twisted sectors of this modding out. The fixed point
subspace is to be identified with the worldbrane of a D-brane.
The operation just sketched shows how one can obtain open strings
by modding out closed  string theories by identifying a  \Z2
transformation in the target space, with a worldsheet orientation
reversal.  Even if we are not discussing
the unoriented string theories, and in particular wish to
have orientable strings,
this suffices to show that we can
consistently couple D-branes to closed
orientable strings.  It also
shows that the concept of D-branes is far more general than having
a flat worldvolume.  Consider an arbitrary string compactification
and suppose it has a \Z2 symmetry with a fixed curved subspace $F$.
Then considering open strings whose boundary lies in $F$ gives
a D-brane which preserves half the supersymmetries (left
intact by the compactification).  Note that for \Z2
being orientation preserving in
the target space, we would be talking about D-branes
of type IIB and for orientation reversal, we would be talking
about D-branes for type IIA strings.
For example consider a Calabi-Yau hypersurface of dimension $n$, which
is defined by an equation with {\it real} coefficients in $\Pp^{n+1}$.
  Then it admits a Ricci-flat
metric which is invariant under the complex conjugation operation.
Moreover the fixed point of the operation of complex conjugation
 can be identified with a
real hypersurface in ${\bf RP}^{n+1}$ and thus defines a
D-brane worldvolume consistent with preserving half the supersymmetries.
A special case of the above that we will use in this paper is
when a Calabi-Yau has a
conifold type singularity which locally looks like
\eqn\conif{z_1^2+...+z_n^2=\mu}
Let us take $\mu$ to be real.
Then the corresponding D-brane lies on the real sphere $S^n$.

{}From the above description one may think that we need
 Calabi-Yau manifolds with \Z2 isometries to have
non-trivial supersymmetric D-branes.  This is not the case.
The simplest way to see this is to start with a D-brane
which comes from an orientifolding description and then deform
the theory by marginal perturbations.  These perturbations will interact
in a supersymmetric way with the degrees of freedom
on the D-brane and modify their shape precisely as to preserve
half the supersymmetries.  It is natural to speculate, given
the simplicity of obtaining supersymmetric
D-branes as deformations of orientifold planes, whether {\it all}
of them can be viewed as obtained from such
deformations.

Given the fact that the D-branes must preserve
half the supersymmetries, and given the fact that their curved
worldvolume does not generally support a covariantly constant
spinor (as the above conifold example shows) we learn that
their supersymmetry cannot be realized in the standard form--
rather they must necessarily involve a twisted definition of
supercharges.
This means that the worldbrane theory is at least a partially
twisted topological field theory!  The partial twisting
should make supersymmetry possible
on the {\it curved} worldbrane, but lead to an ordinary
(perhaps reduced) supersymmetry on the {\it flat} parts
of the worldbrane, similar to what was considered in \bjsv .
 This is quite surprising because
we did not artificially `twist' the definition of supercharge.
Nevertheless the consistent interaction of the D-brane with the
rest of the string modes automatically does that for us!  Had
we studied the D-brane theory before introduction
of topological field theories, we would have been {\it forced}
to discover it by studying the realization of supersymmetry
in the new
subsector of string theory in the presence of D-branes.
Thus topological field theories find their most natural interpretation
in the setup of D-branes in string theory.  In fact this also
suggests that as long as the non-compact directions of a
D-brane worldvolume has dimension less than or equal to $2$,
topological twisting in the rest of the dimensions is responsible
for making sense of an otherwise meaningless field theory.

To understand this point about twisting and the unconventional
Lorentz assignments to fields let us concentrate on the transformation
properties of the scalars.  We will also introduce a notation
which we will use in the rest of the paper.  Consider a
type II string
``compactification'' given by $M^9$.  We are using the word
``compactification'' to include ${\bf R}^k$ as part of $M^9$ for
some $k$.  Let $W^p\subset M^9$ be a $p$-dimensional
spatial part of a p-brane.  Then the bosonic degrees of
freedom, living on the $p+1$ worldbrane $W^p\times {\bf R}$,
consists of a gauge field $A$ on it, plus $9-p$ scalars.
As pointed out in \witb\ changing the expectation
values of scalars can be viewed as moving the D-brane
in the ambient space.   Clearly there are $9-p$ such directions
which can be viewed as the `light cone' degrees of freedom
of the D-brane.
However from this geometric description it is clear that
they cannot be viewed as {\it functions} on $W^p$.  Let
us decompose the tangent bundle of $M$ on the subspace
$W^p$ as
\eqn\tand{TM=TW\oplus N_W}
where $N_W$ is the normal bundle on $W$.  Note that the dimension
of $N_W$ is $9-p$ and it is the correct number of degrees of freedom
to describe the variation of $W$ within $M$.  Thus the $9-p$
``scalars'' are actually sections of this bundle.  Twisting
is forced on us by the geometry of the embedding.
These scalars, in addition to transforming non-trivially
under the holonomy group in $W^p\times{\bf R}$, may be partly
compact.  This is clear when we recall that they describe
the movement of the D-brane in $M^9$ and $M^9$ may be partially
compact.  This will prove to be important  for us to take into account
properly.

We will be interested in counting degeneracy of D-brane
states.  Suppose we
have a type II string compactified on
$$M^9={\bf R}^{9-d}\times X^{d}$$
where $X$ is a compact $d$-dimensional manifold.
To get a particle in the effective  ${\bf R}^{10-d}$
spacetime, we have to wrap all the spatial components of the p-brane
around a compact $p$-cycle $W^p\subset X$.  We will
consider the general situation of $N$ wrapped p-branes and look
for possible bound states.
The basic idea is the observation in \witb\ that,
modulo the complications of bound states at threshold, the
existence of bound
states of D-branes is equivalent to having
a mass gap in the corresponding $SU(N)$ QFT on the D-brane worldvolume
(the $U(1)\subset U(N)$ part of it describes the center of mass degree of
freedom).
Moreover, if there is a gap, the degeneracy of the bound states
is the same as the number of ground states in the corresponding
QFT.
 However if
we are looking for bound states
of $N$  such p-branes
at best we could have a bound state of the
p-branes at threshold, because energetically the BPS bound state
will lead to the same energy as $N$ static and separated
wrapped p-branes.

To avoid the delicate issue of bound states at threshold
it was suggested in \senone\ that if we
indeed do have a bound state at threshold of $N$ p-branes,
by considering
a further compactification on a circle $S^1$, and considering
 a state of the BPS state with momentum $M$ around
$S^1$, and if $M$ and $N$ are relatively prime, there
would be no BPS subsystems with the same energy.  In fact
it is convenient to do a $T$-duality on the circle $S^1$
which maps the momentum mode to a winding mode and at
the same time both exchange type IIA and type IIB strings and change
 p-branes to  (p+1)-branes  incorporating the circle as a part
of the worldvolume of the D-brane.  The existence of a winding
mode of $M$ units means,  through the coupling of worldsheet
$B$-field to the D-brane $U(N)$ gauge field \witb\ that we are turning on
$U(1)$ electrical flux along $S^1$ which translates to
a ${\bf Z}_M$ `t Hooft electrical flux in the $SU(N)\subset U(N)$.
 It is convenient
to take the circle to be infinitely long, in which case
 we should have a mass gap in the corresponding
field theory on the D-brane worldvolume ${\bf R}^2\times \Sigma^p$.
{}From now on we will take for granted that we may talk about
type II A(B)
$p$-branes for $p$ even (odd)
wrapped around $X$ or by the above transformation
consider type IIB(A) compactification on $X\times S^1$ where now
we have a (p+1)-brane
wrapped around
$X\times S^1$.  The reader should be aware of this transformation
throughout this paper as we will be using either formulation
depending on what is convenient for the question at hand.  Note
also we do not need to specify whether we are dealing with type IIA
or B, because that is already implied by the parity
of p for the p-branes under consideration.

We will also briefly consider the notion of D-instantons,
where time is not part of the worldvolume of the D-brane.
In other words, we also consider the case where
the $p+1$ worldbrane is purely Euclidean.  In such cases
we would be mainly interested in compact worldbranes.
These would be configurations relevant for non-perturbative
corrections to string theory \ref\polcom{J. Polchinski, Phys. Rev,
{\bf D50}
(1994) 6041}.

\subsec{Supersymmetric Cycles}
In this section we will discuss the
conditions for having supersymmetric cycles.
A supersymmetric cycle is defined by the  condition that a worldvolume
 theory is supersymmetric \ref\old{J. Hughes, J. Liu and J.
Polchinski,
Phys. Lett. {\bf B180}
(1986) 370}\ref\sist{K. Becker, M. Becker and A. Strominger,
{\it Fivebranes, membranes and Non-Perturbative StringTheory},
Preprint NSF-ITP-95-62, hep-th/9507158}.
The (p+1)-cycle is supersymmetric if the global supersymmetry
transformation
can be undone by
$\kappa$-transformation\foot{This should be
the origin of topological twisting discussed in the previous
section.}, which implies that
\eqn\susycy{P_{-}  \Psi={1 \over 2} \Big(1-{i \over (p+1)!} h^{-1/2}
\epsilon^{\alpha_1...\alpha_{p+1}}
\partial_{\alpha_1}X^{m_1}...\partial_{\alpha_{p+1}}X^{m_{p+1}}
\Gamma_{m_1...m_{p+1}} \Big) \Psi=0~,}
where $\Psi$ is a covariantly constant ten dimensional spinor, $h_{\alpha
\beta}$
induced
metric on p-brane and $\Gamma_{m_1....m_{p+1}}$ are ten-dimensional
$\Gamma$-matrices.
Supersymmetric 2- and 3-cycles on Calabi-Yau
threefolds were considered in \sist.
Supersymmetric 2-cycles are given by {\it holomorphic} curves. The
condition on
3-cycles is more complicated. It requires that the pullback of the
K\"aher form $\omega$
vanishes and the 3-volume is proportional
to the pullback of the holomorphic 3-form $\Omega$,
namely
\eqn\cycy{\ast \Phi(\omega)=0~,~~~\ast \Phi(\Omega)~\sim 1}
where $\Phi(.)$ denotes the pullback and $\ast$ is a Hodge dual on
membrane
world-volume.
 The cycles satisfying \susycy\ minimize the
p-brane volume \sist.

It is a simple exercise\foot{
The
moduli space
of hyper-K\"ahler
structures coincides with the space of 3-planes in ${\bf
R}^{3,19}$.  This
3-plane is spanned  by three
hyper-K\"ahler forms $\{J^i\}$ satisfying the quaternionic
algebra.
The condition \susycy\ can be rewritten as
$$(1-*\Phi(\sigma_i\, J^i))\Psi =0,$$
where $\{\sigma_i\}$ are the three Pauli matrices.
}
to find solutions of \susycy\ for supersymmetric 2-cycles on K3.
One finds that for a given Ricci-flat metric on $K3$
 the supersymmetric 2-cycles
are holomorphic curves with respect to one of the complex structures
compatible with metric. A family of such complex structures is parametrized by
points on \P1. It is easy to prove that for each element $e\in H_*(K3, {\bf
Z})$ with self-intersection $e^2\geq -2$ there is a choice of complex structure
on $K3$ for which it can be realized by a holomorphic curve of genus $e^2/2+1$.
 From now on when we talk of supersymmetric 2-cycles on $K3$ as holomorphic
curves we implicitly assume the appropriate complex structure is chosen.

We will also be interested in supersymmetric cycles
for compactification on $G_2$ and $Spin(7)$ manifolds.
 These manifolds have been recently
studied by  Joyce \ref\joyce{D. D. Joyce, {\it Compact Riemannian
$7$-manifolds with
Holonomy $G_2$: I,II}, Oxford preprint; to appear in
J. Diff. Geo. \semi {\it Compact Riemannian 8-manifolds with Holonomy
$Spin(7)$}, Oxford preprint; to appear in Inv. Math.}\ who constructed the
first
compact manifolds of these types.  They have also been studied
in conformal theory \ref\shatv{S. Shatashvili and C. Vafa, Selec.
Math. {\bf 1}
(1995) 347}.
A $G_2$ holonomy manifold has a canonical 3-form $\phi$.  There are
two natural classes of cycles to consider,
which have already been considered in the mathematics
literature \ref\hlaw{R. Harvey and H. B.  Lawson, Acta. Math. {\bf
148} (1992)
47}:  3-cycles
on which $\phi$ is the volume form (known as associative
manifolds) and four dimensional submanifolds for which $*\phi$
is the volume form (coassociative manifolds).
   For $Spin(7)$ manifold there is a canonical
4-form $\Phi$.  A natural class of 4-cycles correspond to
``Cayley'' submanifolds, where $\Phi$ is the volume form of the
4-manifold.
In each of these cases they minimize the volume in the corresponding
class of the cycle and it is easy
to see that they satisfy \susycy\ and thus correspond to supersymmetric
cycles. Examples of all such cycles were constructed in \joyce\
using fixed points of \Z2 isometries (just as our discussion
above would have led us to search for).

\newsec{Type II Strings on $K3$ and $T^4$}
In this section we shall discuss aspects of 2-brane states
for compactification of type II strings on $K3$ and on $T^4$.
In the next section
we shall connect this to bound states of 0-branes and a 4-brane.

\subsec{2-branes on $K3$}
We consider 2-branes whose spatial degrees of freedom
correspond to $W^2=\Sigma $ where $\Sigma$ is a Riemann
surface of genus $g$ embedded ``supersymmetrically'' in $K3$,
which as discussed before means holomorphically embedded.  Suppose
we have $N$ coincident 2-branes.  As usual we do a duality transformation
by introducing an extra circle, changing
2-branes to 3-branes and ask what is the corresponding field theory
on the 4-dimensional worldvolume of the 3-brane.
Were it not for the curvature in $\Sigma $
the field theory living on the 3-brane worldvolume
 would have been the reduction of $N=1$ Yang-Mills theory with $U(N)$
gauge symmetry
from 10 dimensions down to 4, i.e. an N=4 Yang-Mills theory
in four dimensions.  However as discussed in the previous
section we must have a partially twisted version of $N=4$
theory.  The easiest way to find which version is to consider
the transformation properties of the scalars. As discussed
in section 2, they come from the normal bundle $N_W$
which in this case is 6 dimensional.  The normal
bundle has two pieces. One comes from fixing the 4 coordinates
on ${\bf R}^4$ which  gives 4 real scalars as the normal bundle
(i.e. a position in ${\bf R}^4$).  The other 2 real or 1 complex component
 comes from the normal
bundle $N_\Sigma$
of $\Sigma$ in $K3$. Let us decompose the complex rank 2 tangent bundle
${\cal T}$
of $K3$
along $\Sigma$ as $K^{-1}_\Sigma \oplus N_\Sigma$ where $K^{-1}_\Sigma$ is
the tangent bundle to $\Sigma$. Then take  the determinant,
${\rm det\,}{\cal T}|_\Sigma=K^{-1}_\Sigma\, N_\Sigma$. Since the canonical
bundle
${\rm det\,}{\cal T}$ of $K3$ is trivial,
we learn that the normal bundle
is the cotangent
bundle on $\Sigma $:  $N_\Sigma=K_\Sigma$
 and so we can think of the corresponding complex scalar
to be a 1-form on $\Sigma$.
This is precisely the same partial topological twist of $N=4$ along
$\Sigma$
considered in \bjsv.  Let us denote the scalar fields which are
1-forms on $\Sigma$ by $\Phi_z$ and $\overline{\Phi_{\bar{z}}}$.
  Then the result
of \bjsv\ implies that to find the light degrees of freedom of the gauge
theory on ${\bf R}^1\times S^1$, we first have to solve the
Hitchin equation
$$F_{z\overline z}=[\Phi_z,  \overline{\Phi_{\bar{z}}}]$$
\eqn\hitc{\overline{D}_{\bar{z}}\Phi_z=D_z \overline{\Phi_{\bar{z}}} } where
for $N$ three-branes, the gauge fields and the scalars
are in the adjoint of $U(N)$.  Let ${\cal M}^H$ be
the moduli of solutions to these equations.  Then as a piece
of the effective theory on ${\bf R}\times S^1$ we would get
a supersymmetric sigma model on ${\cal M}^H$.  If the gauge
dynamics on the 2-dimensional subspace can be integrated out,
as would be the case for turning on `t Hooft magnetic
flux on $\Sigma$ where the gauge field is completely broken
 or if we have only one threebrane
where the gauge group is $U(1)$, there would be  no additional
degrees of freedom.  Otherwise we will have in addition
a $U(N)$ gauge theory on ${\bf R}\times S^1$ with $N=4$
supersymmetry (i.e. the field content of the reduction
of 4 dimensional $N=2$ supersymmetric YM down to 2 dimensions).
A case of particular interest
is when we have only one threebrane.
Then we have only a $U(1)$
gauge group, and the above equations simplify:  They correspond
to choosing a flat $U(1)$ bundle on $\Sigma$ and choosing
in addition a harmonic 1-form corresponding to solutions of
the scalar $\Phi$.  Recalling the meaning of $\Phi$ as normal
variation of $\Sigma$ in $K3$, it implies that a choice
of the harmonic 1-form is the same as a choice of an infinitesimal
deformation of $\Sigma$ to a nearby holomorphic curve in $K3$.
In fact this description ``compactifies'' the Hitchin equation.
In other words the higher order terms in the effective theory
of the worldbrane retain the information about the meaning of
$\Phi$ as a normal variation of $\Sigma$ in $K3$. Thus
the actual moduli space we  end up with will be a kind of
``compactification'' of ${\cal M}^H$.  What this means concretely
for us is that this space is the same as the moduli of
holomorphic curves of genus $g$ with a choice of a
flat $U(1)$ bundle, i.e. a point on the
Jacobian of the Riemann surface. Let us denote this space for the
case of genus $g$ curve by ${\cal M}^H_g$.

This space can be  studied mathematically, see for example
\ref\Donag{R.~Donagi,
L.~Ein and R.~Lazarsfeld, {\it A non-linear
deformation of
the Hitchin Dynamical System}, preprint 1995}. By definition, ${\cal
M}^H_g$ is fibered over the space of holomorphic genus $g$ curves
on $K3$. The
space of such curves can be identified as $\Pp H^0({\cal
O}(D))=\Pp^g$, where
$H^0({\cal O}(D)$ consists of functions with a simple pole on the
divisor $D$
represented by curve of genus $g$. The fiber of ${\cal M}^H_g$ over
a point on
$\Pp^g$ is a Jacobian of the corresponding curve.
 This space has complex dimension $2g$, it is hyperk\"ahler and has the
structure of a completely integrable system.

Let us consider the two simplest examples. For $g=0$, the holomorphic
(rational) curves on $K3$ are isolated. Moreover the Jacobian for
$g=0$ is
trivial, so
${\cal M}^H_0$ is a point.  Equivalently, there is no non-trivial
solution to \hitc.  This is also true even if we consider $U(N)$
instead of $U(1)$, a fact we shall use later.

For $g=1$, the space of elliptic curves in $K3$ is \P1. The
Jacobian of the
elliptic curve coincides with the curve itself. Generically the
curve is smooth,
but there are 24 degenerate fibers which are rational curves with a
double
point. Such fibration over \P1 fits together to form a $K3$!
Thus ${\cal M}^H_1=K3$.

For $g>1$, we will now
prove\foot{We thank E.~Witten for the illuminating
 remark which contained the idea of this proof.} that a space of curves on
 $K3$ with line bundles of degree $g$  is  birationaly equivalent to a
symmetric product
${\rm Sym}^{g}(K3)$.
A point on ${\cal M}^H_g$ may be viewed as a
curve in $K3$ together with $g$ unordered points
on $\Sigma$ (we use an isomorphism
${\rm Jac}_g={\rm Sym}^g(\Sigma)$). Forgetting the curve gives $g$ points in
$K3$, i.e.~a point of ${\rm Sym}^{g}(K3)$. Now fix a point on $K3$. All curves
 of genus $g$ passing through this point are parameterized by a hyperplane in
$\Pp^g$. For $g$ points on $K3$ one gets $g$ hyperplanes in $\Pp^g$, which
always intersect, genericaly in a point. Thus for any $g$ points on K3 there
is at least one (genericaly one) curve of genus $g$ passing through them.
Tautologicaly this curve has $g$ marked points on it\foot{
This is very much like the $N=2$ strings on $K3$ \ref\oova{
H. Ooguri and C. Vafa, {\it All Loop $N=2$ String
Amplitudes}, hep-th/9505183}\ where
the $N=2$ string computes the `number' of holomorphic maps
from genus $g$ Riemann surface to $K3$.  In fact in the context
of type IIB theories and Euclidean 1-branes, the analogy may become
more exact.  Note in particular that $N=2$ string also has a $U(1)$
gauge bundle on the worldsheet, as is required for the D-brane.  This
suggests that $N=2$ string
(and perhaps other topological strings as well)
may be viewed as the theory of appropriately embedded
Dirichlet 1-branes.}. This proves the
birational equivalence.  It is easy to see that the map is symplectic.

This theorem is not quite what we need. It may be viewed as a mathematical
motivation for a  conjecture that  ${\cal M}^H_g$ is actualy related to
${\rm Sym}^{g}(K3)$ and has the same cohomology in particular. A proof of
this conjecture would involve analysis of singularities of both varieties
using  a theorem due to Lazarsfeld \ref\lazar{R.~Lazarsfeld, {\it
J.~Diff.~Geom.}~{\bf 23} (1986)
299-307}.

Note that the number of D-brane states is the same as the
dimension of the ground state for the worldbrane theory, which
for one threebrane case is the same as the number of cohomology
elements of
${\cal M}^H_g$.
Type II--heterotic duality in 6 dimensions predicts this number
to be the same as the degeneracy of the bosonic oscillation
states with $N_L-1={1\over 2}P^2$ where $P^2$ denotes the
self-intersection
of the 2-brane part of the 3-brane in $K3$, i.e. $P^2=\Sigma \cdot
\Sigma=
2g-2$.  In other words we expect a degeneracy of bosonic
oscillators for level
$N_L=g$.  This is in accord with what we mentioned for $g=0$, where the
sigma model is trivial,
and $g=1$, where the sigma model is the one on $K3$ and so has dimension
24.  In addition
the $U(1)$ center of mass system gives rise
to the expected supersymmetry degeneracy.  Moreover for the higher values of
$g$ the mathematical conjecture about
the form of ${\cal M}^H_g$
 is in accord with the
suggestion made in \va\ as to how the degeneracy
of Hagedorn density of BPS states may arise.  In fact we will see
in the next section, after uncovering an anomaly,
 how the counting proposed in
\va\
for $n$ 0-branes and one 4-brane gives the answer
predicted by string duality.  Thus the $T$-duality
of string theory, which maps the 0-brane--4-brane configuration
to 2-brane configuration,
 would show
the sensibility of the above mathematical conjecture about the
structure of ${\cal M}^H_g$.

The case when $\Sigma$ is a 2-sphere is special.  This
is the case when the cycle can vanish by varying the moduli
of $K3$, and  so could lead to massless particles.  One of the
predictions of string duality is that when this happens for a
D-brane wrapped around the 2-sphere we have exactly one BPS state,
which is what we have found above, namely ${\cal M}^H_0$ is a point.
There is a further prediction: If we take the D-branes to wrap
around the 2-sphere $N$ times, or in other words, if we have
$N$ coincident 2-branes, they should not form a bound state, otherwise
we would end up with more massless modes than is expected
on the heterotic side. The mathematical counterpart of this statement
is quite clear: A cycle which is $N>1$ times 2-sphere has
self-intersection
number equal to $-2N^2<-2$. Hence it cannot be realized by a smooth
holomorphic
curve, or a supersymmetric 2-cycle.
We will now confirm that this is indeed the
case.

As mentioned above, when we have $N$ D-branes, we have a $U(N)$
gauge group to deal with.  Since the Hitchin equation in the case
of the $S^2$ has a trivial solution, the gauge symmetry is unbroken
and so in the remaining 2 dimensional theory we obtain an $N=4$
$U(N)$ gauge theory (with no extra matter).  Modding out the
center of mass degree of freedom, we will concentrate
on the $SU(N)$ gauge symmetry and as discussed in section 2
we will consider subsectors with ${\bf Z}_M$ electrical flux and look
for mass gap.  The absence of mass gap will show that we will not
have a bound state (and in fact implies that supersymmetry
is broken in this subsector).  As noted in \witb\ the usual
anomaly matching conditions of `t Hooft imply that
for $N=4$ YM in $d=2$,
unlike the $N=8$ case, there is no mass gap (to see that note
that the $R$ symmetry group is $O(4)$ and the left-moving
fermions can be chosen to transform as $(2,1)$ and right-moving
ones as $(1,2)$.  Non-triviality of
the anomaly matching in any of the $SU(2)\subset
O(4)$ implies the lack of mass gap in this theory).  This implies
that, as expected from
string duality,
$N$ 2-branes wrapped around $S^2$ do not form a bound state.

\subsec{Bound states and $U(n)$ Hitchin systems.}

Here we would like to discuss the situation with the bound states
corresponding to supersymmetric 2-cycles in $X$, where $X$ is either
$K3$ or
$T^4$.
It was already explained that the rational cycles, which have
self-intersection
$-2$, do not form bound states. The situation is however different
for the
cycles of genus $g>1$. The point is that a  class $e\in H_*(X,{\bf Z})$
which is  a
multiple of another class $e=n\cdot e',\, n\in {\bf Z}_+$ can be
represented
either by a smooth curve of genus $g$ or an n-tuple of
coinciding
curves of genus $g',\, g=n^2(g'-1)+1$. The second representation
allows us to
think of such D-branes as bound states of
simpler incident D-branes. Since this
describes the   same physical
object, we should recover the same structure of moduli space using either
description.
This gives yet another interesting test of D-brane picture: The
moduli space of
curves of genus $g$ with $U(1)$ bundles was described above --- it is a
$2g$--dimensional hyperk\"ahler variety. On the other hand,
following Witten,
the moduli of n-tuples  of curves of genus $g'$ can be described as
follows. A
point on the moduli is a pair --- a complex rank $n$ vector bundle $E$ on
$\Sigma$ together with $n^2$ ``scalars'' $\phi^{ij}_z$. These
``scalars'' are
sections of
${\rm End}\,{\cal E}\times N_\Sigma$. They correspond to a condensate of
D-strings connecting pairs of 2-branes.  Such moduli space is indeed a
hyperk\"ahler variety
of requisite dimension $2g$! Locally one can think of it as a set of
solutions
of $U(n)$ Hitchin's system on curve of genus $g'$.
A similar problem was studied in some detail in \Donag.
There it is shown  that indeed there is an isomorphism between the
moduli of
curves of genus $g$ with $U(1)$ and the moduli of n-tuples of
curves of genus
$g'$ --- a certain compactification of $U(n)$ Hitchin system.

\subsec{Type II on $T^4$}
The situation for type II compactification on $T^4$
(or $T^4\times S^1$) and the
resulting 2(3)-branes is very similar to our discussion
for the $K3$ case, so here we highlight the differences.

 It is not difficult to see using standard techniques that the
moduli space of holomorphic
curves of genus $g>1$ in $T^4$ is a product $\Pp^{g-2}\times
T^4$. (Obviously, the $T^4$ factor accounts for the action of torus on
itself.)
If we
consider curves together with  flat $U(1)$ bundles, we get a
hyperk\"ahler
variety of dimension $2g$ fibered over $\Pp^{g-2}\times T^4$ by Jacobians
$J_g=T^{2g}$. For $g=1$ the space of elliptic curves in $T^4$ is
$T^2$ and the
space of elliptic curves with Jacobians is $T^4$. From above it
follows that\foot{The result is in agreement with the result
of \sentwo\ where two tori intersect at one point
and lead to a degenerate genus 2 curve.}
the moduli space for $g=2$ is $T^4\times T^4$. Clearly, two factors of
$T^4$ correspond to the action of $T^4$ on the curve itself and on its
Jacobian. In general, for $g\geq 2$ the moduli space have a structure
$T^4\times X$ where in turn $X$ is a fibered product of $T^4$ over a
$(2g-4)$-dimensional hyperk\"ahler manifold $Y$. Finally,
$Y$ has a structure of a family of complex $(g-2)$-dimensional tori over
$\Pp^{g-2}$.

For $g=3$, $Y$ is a hyperk\"ahler 2-fold. Therefore, assuming compactness,
this is either a torus
$T^4$ or a
$K3$. Since it is
fibered over \P1, it has to be $K3$. Thus the moduli space for $g=3$ is a
product of
$T^4$ and a $T^4$ bundle over $K3$. In the next section we shall compare this
answer with what is expected from string duality.

\newsec{Gas of 0-branes on $K3$}
It was shown in \va\ that a gas of $n$ 0-branes on $T^4$
in the presence of one 4-brane gives rise to a sigma
model on $(T^4)^n/S_n$ where $S_n$ is the permutation group
on $n$ objects. Moreover, the ground states of this system
is as many as expected by string duality, i.e. left-moving
superstring degeneracy at level $n$.  This is based on
the observation in
\ref\vw{C. Vafa and E. Witten, Nucl. Phys. {\bf B431} (1994) 3}\
about the
relation of the cohomology of this space
to string oscillator partition function.

Let us briefly review the basic conclusion
of \va , before
considering the  more involved
case of $K3$.
By T-duality for $T^4$, a supersymmetric  2-cycle of genus $g$,
with $g-1\geq
0$,
can be mapped into a sum of a fundamental 4-brane $[T^4]$ and $(g-1)$
0-branes represented by points on $T^4$. The configuration
 space of such a system, taking into account that
one  0-brane and one 4-brane form precisely one bound state
\sentwo\ is ${\rm Sym}^{g-1}(T^4)$ coming
from the position of the 0-brane, and  a $T^4$
form the 4-brane (due to
the choice of a flat
$U(1)$ background for the 4-brane). So finally
the moduli space
of the D-brane configuration we obtain is
a product $T^4\times {\rm Sym}^{g-1}(T^4)$ in agreement
with the prediction of $U$-duality \va.
 This should reproduce
the answer
for the moduli space of 2-branes we obtained above!

This seems to be very likely, in view of the following facts. Both
spaces are
hyperk\"ahler, of the same dimension $2g$. Both have the product structure
$T^4\times X$ where in turn $X$ is a $T^4$ bundle over a $(2g-4)$-dimensional
hyperk\"ahler manifold $Y$.
(This was explained above for the moduli of 2-branes. This is so here
because $T^4$
acts freely on the symmetric product ${\rm Sym}^{g-1}(T^4)$ and $Y$ is defined
as a quotient of this action.)
Moreover, we can
study explicitly three examples.

For zero (one) 0-brane on $T^4$ the answer is $T^4$ ($T^4\times
T^4$) which
agrees with what we had for the  curves of genus one (two) on $T^4$.
Let us see how the answer for two 0-branes on $T^4$ agrees with that for
genus three
curves. In this case ${\rm Sym}^{2}(T^4)$
has a
singular locus on the diagonal $T^4$, with $A_1$ singularity in the
transversal direction. We can blow up along the diagonal which amounts to
creation of a smooth divisor \P1 $\times T^4$.
There is a natural free action of $T^4$ on $T^4\times T^4$: $(a,b)\to
(a+\xi,b+\xi)$. Symmetric group $S_2={\bf Z}_2$ acts by $(a,b)\to (b,a)$.
Let us consider the quotient $(T^4\times T^4)/T^4$. Obviously, this quotient
 is the 4-torus $T^4$ again, so one can think of $T^4\times T^4$ as a
(nontrivial) fibered product of $T^4$ over $T^4$. An advantage of this point of
view is immediate:
the diagonal $T^4$ acts
only on the fiber and, more importantly, ${\bf Z}_2$ acts only on the base.
Since
$T^4/{\bf Z}_2=K3$ we proved that $Y=K3$ so that ${\rm Sym}^{2}(T^4)$ is a
$T^4$ bundle
 over $K3$. This is
exactly the result we obtained above for the genus 3 supersymmetric
2-cycles!  For higher genus, the arguments given in the context
of 2-branes on $K3$ is presumably
generalizable to the case of $T^4$.

It was suggested in \va\ that the same
idea should work in the context of
bound states of 0-branes and one 4-brane
on $K3$, in a way consistent with type II--heterotic duality,
when we take into account the observation of \vw\ that
$n$ copies of $K3$ has the same degeneracy as bosonic
string partition function at level $n$.  In fact
the gauge system in this problem is identical to that encountered
in \va\ in connection with the $T^4$
compactification,  except for lacking
the adjoint matter which came from the reduction of the 4-brane
on $K3$.  In the case of $T^4$ this system was decoupled and its
fermionic degrees of freedom provided some of the degeneracies
expected for the BPS state.  In our case they are absent,
again in accord with the fact that we have a different degeneracy
of BPS states expected.  Everything works as in the $T^4$
case, except that when we have $m$ 0-branes  and one 4-brane,
the corresponding sigma model we get is on ${\rm Sym}^m(K3)$,
which has the same ground state degeneracy
as the bosonic string with $N_L=m$.
However the string duality in this case predicts that
$N_L-1=m$ since ${1\over 2}P^2=m\cdot 1=m$.  Thus there
seems to be a shift of one unit unaccounted for.

Here we will argue that in the presence of a $K3$ a 5-brane
acquires in addition -1 unit of the RR charge corresponding
to the two form.  This shifts the ${1\over 2}P^2\rightarrow
(m-1)\cdot 1=m-1$, which thus leads to the observed degeneracy
in agreement with $N_L=m$.  We will argue for this by making
use of type IIB $SL(2,{\bf Z}) $ duality in 10 dimensions,
though it should also be possible to check it directly
in string perturbation theory.  Consider a 5-brane.
We will argue that in the 6d worldvolume of the 5-brane
there is an effective interaction of the form
\eqn\anam{{1\over 48}\int_{M^6} B^R\wedge p_1}
where $B^R$ is the RR 2-form gauge field
and is normalized to have periodicity $2\pi$
and $p_1$ is the first Pontryagin class of the manifold.  We will not
be careful with signs, but this result shows that
if we compactify the 5-brane on a $K3$ where $p_1=-48$,
 in the resulting 2d
theory, we have an effective one unit of RR 2-form charge.  In
other words
a 5-brane compactified on a $K3$ acquires an additional charge
as if there were an anti-1-brane on $K3$.  We will show that the
absolute value of the shift is one, but as in \ref\vwo{C. Vafa and
E. Witten,
Nucl. Phys. {\bf B447} (1995) 261-270}\ we will  have to confess to not
checking the sign of the additional
charge.

The most straight forward way to check \anam\ would be
to compute a three point function of the RR 2-form field
with emission of two gravitons on a disc  where
the boundary of the disc is mapped to the worldvolume of the
5-brane.  We have not done this, but have found an indirect argument
for its existence by making use of type IIB $SL(2,{\bf Z})$
duality in 10 dimensions.  Consider type IIB strings
in the presence of $n$ parallel fivebranes.  Then according
to the prediction of $SL(2,Z)$ duality, by strong/weak
duality this is equivalent to $D$-string (the (0,1) string
in the notation of \sch) in the presence of
symmetric fivebrane of charge $n$, which according
to the result of \ov\ is equivalent by T-duality to
type IIA on $A_{n-1}$ ALE space.  Now we recall that if we compactify
type IIA on $K3$ we induce a one-loop correction of the form
\vwo
$${1\over 2} \int_{M^6}B^{NS}\wedge p_1$$
where $B^{NS}$ denotes the 2-form gauge field from the
NS-NS sector.  It is not too difficult to generalize the
result in \vwo\ to the case of ALE space $A_{n-1}$ where one finds
a quantum correction of the form
$${n\over 48}\int_{M^6}B^{NS}\wedge p_1$$
One way to see this is to recall that $K3$ can be viewed
as $24$ cosmic strings \ref\scs{B. R. Greene, A. Shapere, C. Vafa
and S.-T. Yau, Nucl. Phys. {\bf B337} (1990) 1}\
and $n$ cosmic strings make up an
$A_{n-1}$ singularity (see \ov\ for a discussion of this).
To see what this induced correction means in the original type
IIB string theory (i.e. in the (1,0) string) we recall
that under $SL(2,{\bf Z})$ duality we have an exchange
$B^{NS}\leftrightarrow B^R$.  Thus we end up with
the expected result \anam.

\newsec{Calabi-Yau Conifold}

In this section we will briefly mention the situation
with counting of BPS states for 3-branes wrapped around
a vanishing 3-sphere near a Calabi-Yau conifold.
The arguments are very similar to the $S^2$ 2-branes in $K3$
and so we will be rather brief here.  Clearly also other
types of 3-cycles can be studied with the same
techniques as used for higher genus curves in $K3$, but we will not
consider
them in this paper.

In the 3-brane worldvolume we have $S^3\times {\bf R}$.
Let us first note that since near the conifold the Calabi-Yau
is approximated by $T^*S^3$,
 the 6 `scalars' will transform as
a 1-form on $S^3 $ and three scalars.  Thus
if we consider a single 3-brane, there
is a unique vacuum, since there are no
non-trivial flat connections on $S^3$ and moreover
the dimension of the space of harmonic 1-forms is 0
(which implies that the $S^3$ is rigid).  This implies
that we have a single BPS state corresponding to 3-brane
wrapping around the 3-cycle.

Now we show that there are no bound states when we wrap the 3-brane
$N$ times around $S^3$. Again the internal theory will have
a trivial moduli space and we are reduced (modulo the
center of mass degree of freedom) to a reduction of $SU(N)$
$N=1,d=4 $ system down to $d=1$. Again by the
trick of \senone\ to avoid the complications of
bound states at threshold we transform the problem to the
$SU(N)$ gauge system in $d=2$ which has $N=2$.  Again this
system has no mass gap (as noted in \witb)
using `t Hooft's anomaly matching condition.  Thus there are no
multiple bound states of the threebrane wrapped around the $S^3$
of the conifold, in agreement with the proposal in \strom .

\newsec{$G_2$ and $Spin(7)$ Manifolds and $N=4$ Topological Twists}
In this paper we have seen that in the worldvolume of a D-brane
we end up with a twisted version of an $N=4$ theory.  In this
section we show that we can obtain all the three twistings
of $N=4$ by considering different 3-branes embedded in
appropriate manifolds.  We will also point out the existence
of potentially
new massless solitons when we compactify on manifolds of $G_2$
and $Spin(7)$ holonomy when appropriate  3 or 4 cycles vanish.

Let us recall that $N=4$ theory in $d=4$ has three inequivalent
twists \vw\ depending on how we embed the Lorentz group
$SU(2)\times SU(2)$ in the global symmetry of $N=4$ which
is $SU(4)$.  We can classify the embedding by how the fundamental
$\bf{4}$ of $SU(4)$ decomposes.  There are three cases\foot{
There is a misprint in case i) above considered in \vw .} which will lead
to a topological symmetry: i) $({\bf 2},{\bf 1})\oplus ({\bf 1},{\bf 2})$
$ii) ({\bf 1},{\bf 2}) \oplus
({\bf 1},{\bf 2})$ iii)$ ({\bf 1},{\bf 2})\oplus ({\bf 1},
{\bf 1}) \oplus ({\bf 1},{\bf 1})$.  The cases i) and ii) lead
to 2 unbroken topological supersymmetries, whereas the case iii)
leads only to 1 unbroken topological supersymmetry.  Let us consider
the transformation of the `scalars' in each case.  The scalars
transform as the ${\bf 6}$ of $SU(4)$, i.e. the second rank
antisymmetric representation.  Thus under the twisted Lorentz
group the scalars now transform as
$$i) \qquad 2({\bf 1},{\bf 1}) \oplus ({\bf 2},{\bf 2})$$
$$ii) \qquad 3({\bf 1},{\bf 1}) \oplus ({\bf 1},{\bf 3})$$
$$iii) \qquad 2({\bf 1},{\bf 1})\oplus 2({\bf 1},{\bf 2})$$
Let us consider a Euclidean 3-brane, i.e., we look for 4-brane
configurations in compact spaces.  These would serve as instantons
for string theory.  Then the above transformation of the scalars,
combined with our discussions in section 2 imply that
in cases i) and iii)  we should look for an 8-manifold compactification,
whereas in case ii) we should look for a 7-manifold compactification.
Moreover taking into account the number of unbroken topological supersymmetries
in each of these cases, we see that the candidates are fixed to be
i) a Calabi-Yau 4-fold compactification, ii) $G_2$
holonomy 7-manifolds, iii) $Spin(7)$ holonomy 8-manifolds.

Example for case i) is easy to come by --- we have essentially
mentioned it in section 2, where we discussed Calabi-Yau
compactification with complex conjugation symmetry.  The
4-brane in this case will be the real 4-manifold and the normal
bundle will be its cotangent bundle, so that the non-trivial
scalar will be identifiable with 1-forms on it, i.e. transform
as $(2,2)$ as expected for case i).

It turns out for coassociative four submanifolds in $G_2$
the normal bundle is indeed $(1,3)$ \ref\mar{R. C. McLean, {\it
Deformations
and Moduli of
Calibrated Submanifolds}, Ph.D thesis, Duke University, 1990}\
as expected from the discussion
above.  For the $Spin(7)$ case the normal bundle
is of the form $S_+\otimes V$ where $S_+$ is a spin bundle of a given
chirality and $V$ is a two dimensional bundle \ref\joyp{D. Joyce,
private communication.}.
When $V$ is trivial, this
becomes $S_+\oplus S_+$, which is exactly what we had expected for
case iii).

Now we come to the question of D-branes as solitons\foot{In
the case of $Spin(7)$ and Calabi-Yau fourfold
compactifications note that
to cancel the anomaly computed in \vwo\ we need to turn on
some $D$-branes on the manifold to stabilize it.  This should be
interesting to study further.}.
We will be brief because some of the mathematical aspects
of moduli space of compactifications on $G_2$ and $Spin(7)$ manifolds
have not been worked out.  However it is clear that
if we have a vanishing associative, or coassociative or Cayley
submanifolds, we will have a massless state.  What is not
clear yet is whether these will appear at
finite distance in moduli space, and to what extent the moduli
space metric receives worldsheet and quantum corrections.
But clearly once this is known, the methods discussed
above will be easily applicable in counting their bound states.

We would like to thank Andrei Johansen,
Dominic Joyce and Tony Pantev and Edward Witten
for valuable discussions.
The research of  V.~S.~is supported in part by NSF grant DMS 9304580.
The research
of M.~B.~and C.~V.~is supported in part by NSF grant PHY-92-18167.
The research of M. B. was also supported by NSF 1994 NYI award
and DOE 1994 OJI award.

\listrefs

\end